\documentclass[useAMS,usenatbib]{mn2e}

\usepackage{graphicx}
\usepackage{natbib}
\usepackage{amssymb}
\usepackage{lscape}
\title[A variation of the $\Gamma$ within M83]{A variation of the fraction of stars that form in bound clusters within M83}
\author[E.Silva-Villa, A. Adamo and N. Bastian]{
E. Silva-Villa$^{1,2}$\thanks{ES-V: esteban.silvavilla@phy.ulaval.ca}, A. Adamo$^{3}$ and N. Bastian$^{4}$ \\
$^{1}$ Centre de Recherche en Astrophysique du Qu\'ebec (CRAQ) \\   
$^{2}$ Universit\'e Laval. 1045, Avenue de la M\'edecine, G1V 0A6 Qu\'ebec, Canada\\
$^{3}$ Max Planck Institut f\"ur Astronomie, K\"onigstuhl 17 D-69117 Heidelberg, Germany\\
$^{4}$ Astrophysics Research Institute, Liverpool John Moores University, Egerton Wharf, Birkenhead, CH41 1LD, UK\\}
\begin{document}

\date{Accepted 2013 August 09. Received 2013 August 01; in original form 2013 May 13}

\pagerange{\pageref{firstpage}--\pageref{lastpage}} \pubyear{2013}

\maketitle

\label{firstpage}

\begin{abstract}
Recent observations, as well as theoretical studies, have suggested that stellar cluster formation may depend on local and global environmental properties. In particular, the fraction of stars that form within long-lived bound clusters ($\Gamma$) may depend on environment, with indications that it may be higher in the more extreme environments of high star-formation rate density galaxies.  How $\Gamma$ varies has important implications on the use of clusters to determine the star-formation histories of galaxies as well as our understanding of the star-formation process itself.  Previous studies have estimated $\Gamma$ over full galaxies, making it difficult to discern the physical cause of the reported variations. Here, we use existing star cluster catalogues and HST-WFC3 V and I images of the grand design, face-on spiral galaxy M83 in order to see if and how $\Gamma$ varies within a single galaxy.  We find that $\Gamma$ decreases strongly as a function of galactocentric radius, by a factor of $\sim5$ over the inner $\sim6$~kpc, in agreement with recent theoretical predictions and decreasing trends observed in the gas surface density of the galaxy.
\end{abstract}	


\section{Introduction}
During the past decades star clusters have been used as tools to follow the star formation histories (SFHs)
of galaxies at distances where single stars are beyond instrumental resolution. But, how accurately
can star cluster systems trace the real SFH of a galaxy? To understand galaxy formation and evolution it is important to understand how field stars and star clusters 
relate to each other. However, the (actual) amount of star formation happening in star clusters (that
remain bound beyond the embedded phase) is still uncertain. If clusters will be used as tools to trace the 
SFH of a galaxy, it becomes of key importance to understand what fraction of star formation is happening in
long-lived clusters, and whether or not this fraction is correlated with other host galaxy parameters. 
In this paper we will adopt the notation of \citet{bastian08}, and will refer to the fraction of stars that form in clusters (with $\tau\ge10$ Myr) as $\Gamma$.

Various studies in the past decade have attempted to quantify $\Gamma$, each using slightly different techniques. 
\citet{gieles10} estimated that $\Gamma$ varies between 5-18\% for the spiral galaxies M74, M51 and M101, using the observed cluster luminosity function to estimate the cluster formation rate (CFR).  For a sample of five spiral galaxies, \citet{silvavilla11} used resolved stellar populations and star clusters
to estimate $\Gamma$, finding values lower than $\sim15\%$.
\citet{goddard10} studied the star cluster system of the galaxy NGC~3256, and complemented the sample
using results for the CFRs and SFRs from different authors for the LMC, SMC, Milky Way, M83, NGC~6946 and 
NGC~1569. The results found by Goddard et al. suggested a relation between the
$\Gamma$ and the surface star formation rate density ($\Gamma\propto\Sigma_{SFR}^{0.24 \pm 0.04}$). 
\citet{adamo11} studied starburst galaxies (i.e. high SFR) to study the relation of the cluster populations with the host
galaxy, finding relatively large values for $\Gamma$ ($\sim50\%$). \citet{cook12} looked at a sample of local dwarf galaxies with relatively low SFRs and $\Sigma_{SFR}$. As single systems, these galaxies did not seem to follow Goddard's relation, most 
likely because in such low star forming regimes, cluster formation is dominated by stochasticity. 
However, when they considered their galaxy sample as a whole (allowing for a better sampling of cluster formation 
process), the derived  $\Gamma$ is in agreement with the Goddard relation (i.e. $\Gamma$ values of a few percent). 

\citet{kruijssen12} developed a theoretical framework to understand the fraction of star formation that happens in clusters.  Here, bound clusters represent star formation that happens in the high-density tail-end of a hierarchical distribution of the interstellar medium (ISM).  These high density regions attain a high star-formation efficiency, allowing young clusters to survive the expulsion of the residual gas left over from the star formation process.  Based on this framework, \citet{kruijssen12} recovered the $\Gamma$ vs $\Sigma_{SFR}$ relation observed by \citet{goddard10}. A key prediction of the model is that $\Gamma$ should also vary within a single galaxy.

\begin{figure}
\includegraphics[width=\columnwidth]{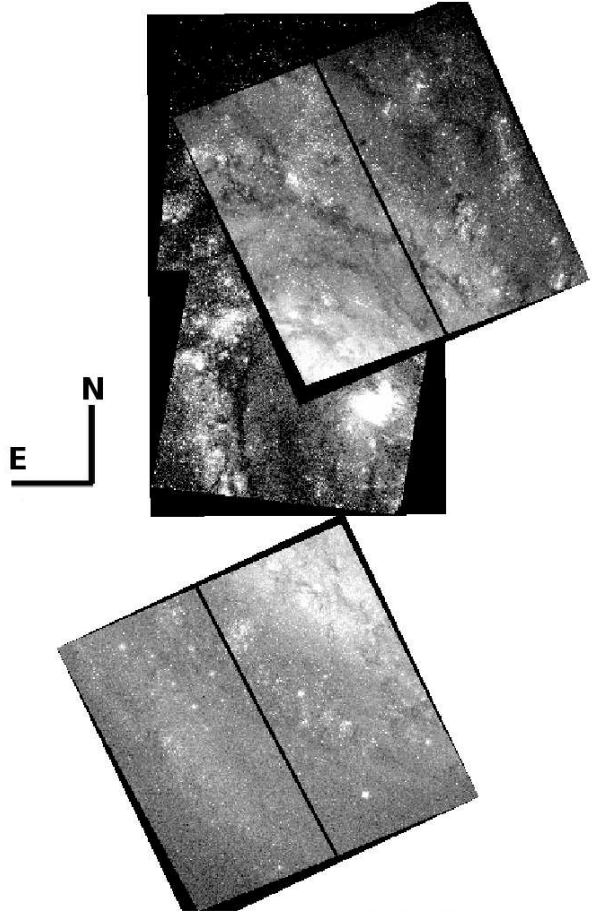}
\caption{WFC3 and ACS fields covering NGC~5236. See text for the details of the observations.}
\label{fig:m83}
\end{figure}

The observed increase of $\Gamma$ in galaxies with increasing intensity of star formation suggests a 
strong relation with the host galaxy. Moreover, there appears to be a tight relation between giant molecular clouds (the birth sites of clusters)
and galactic environment \citep[e.g.][]{hughes13,schinnerer13}, which suggests that the formation and disruption of star 
clusters inside a galaxy must be linked with their ambient environment. However, there has been no study to date on whether $\Gamma$ varies within a galaxy, and if so, by how much.  Using existing {\em Hubble Space Telescope Wide Field Camera 3} (HST-WFC3) star cluster
catalogues and the field stellar population of the galaxy M83, we present the first estimations of the variations of $\Gamma$ inside a galaxy.
We have combined available star cluster catalogues of M83 \citep{silvavilla11,bastian11}.
The sample partially covers the galaxy from north to south (see Fig. \ref{fig:m83}), 
which allows us to study the variations of the star and cluster populations (hence $\Gamma$) at different locations within the galaxy. 
Radial bins, with variable width but same amount of clusters per bin, have been used to avoid size-of-sample effects.
The proximity of M83 \citep[D$\approx4.5$ Mpc,][]{thim03} allows the study of the cluster and field stellar population
with great detail. 


\section[]{Description of the Data}

\subsection{Star cluster catalogues}

For the present study here, we combine the cluster catalogues obtained by \citet{silvavilla11} and \citet{bastian11}. 
The addition of the catalogues from Silva-Villa \& Larsen increase the area coverage. However,
the areas are largely constrained by the fields-of-view of the HST-WFPC2, which in turns does not significantly improve the number of cluster to the total sample.  The inclusion (or exclusion) of the Silva-Villa \& Larsen catalogue does not affect the main conclusion of this work

For a full description on the selection on clusters and their respective age/mass estimations and classification, we refer
the reader to the works of \citet{bastian12} and \citet{silvavilla11}.

Due to the large overlap between ACS and WFC3 fields in the north of the galaxy, the catalogues 
were matched to avoid repetitions, after converting the ACS coordinates into the WFC3
system. For the coordinate transformation we used the task {\em wcsctran} in IRAF.

The final master catalogue contains 1354 star clusters, where 992 clusters have estimates of 
age and mass. The remaining 362 objects were only covered with 3-band photometry (BVI), 
hence estimations of ages and masses for those clusters will be unreliable due to different degeneracies (e.g., age-metallicity).

As part of a larger study (Silva-Villa et al. in prep.) we have constrained our sample as follow: Ages
between 10 and $\sim$300 Myr ($7.0 \le Log(\tau/yr) \le 8.5$), masses above $5 \times 10^3$ $M_{\odot}$, magnitudes
brighter than $M_V \le 22.5$ and clusters with galacto-centric radii equal or larger than 370 pc. 
The limits were chosen to avoid classification confusion for young sources (i.e younger than 10 Myr), 
minimise the effects of stochasticity from the random sampling of the stellar initial mass function for low-mass clusters (i.e. larger than few times $10^3$ M$_{\odot}$), 
incompleteness for older sources (i.e. older than $\sim300$ Myr), and to avoid the centre of the galaxy where
the extinction is highly variable, causing a highly spatially dependent completeness limit.
A total of 429 objects (out of the 992) passed our criteria. The majority of the clusters come from Bastian's catalogue,
due to the poor coverage at the south of the galaxy, with only a very small fraction of clusters passing our criteria. The
reason for the low number of clusters in the south regions come from the small area covered by the HST-WFPC2 camera,
which largely constrains the number of clusters with U-band photometry, required for deriving cluster ages and masses.

The cluster population of M83 has been widely studied. Recently 
\citet{mora09,chandar10,silvavilla11,bastian11, fouesneau12} and \citet{bastian12}
have used HST (WFC3, ACS and WFPC2) imaging to study the cluster population
of this galaxy, with some of the observed fields spatially overlapping. \citet{chandar10}, \citet{whitmore11} and \citet{fouesneau12} studied the inner region of M83 using the same WFC3 imaging as \citet{bastian11, bastian12}. 
Bastian et al. made a comparison between their catalogue and the one
obtained by \citet{chandar10} (for the region contained in both studies), and found that the main differences were due to 
the classification between clusters and associations, which is largely confinded to young objects ($< 10$~Myr).
Above 10~Myr, the catalogues were in reasonable agreement, with only small differences.  Since we limit our analysis to clusters with ages above 10~Myr (see \S~3.1.1) our results are independent of the catalogue used.

\subsection{Field stellar photometry}
We performed point-spread function (PSF) photometry and summarize our procedures here. With a set of bona-fide stars visually selected in our images
(measuring their FWHM with {\em imexam}, with typical values of $\sim2.4$ pixels $\sim2.1$ pc), we constructed
an empirical PSF using the {\em PSF} task in DAOPHOT. The PSF stars were selected
individually in each band, in order to appear bright and isolated. PSF photometry was done with DAOPHOT
in IRAF. This procedure was applied in the same manner for each band (i.e. V and I). 

Our PSF-fitting magnitudes were corrected to a nominal aperture radius of 0\farcs{5}, following standard procedures.
HST zeropoints were applied to the PSF magnitudes after applying aperture corrections. Zeropoints were taken from the HST webpage.
Foreground extinctions values were taken from {\em Nasa Extragalactic Database} (NED).

\section{Radial distribution}

\subsection{Star cluster population}
After converting all coordinates from ACS/WFPC2 into the WFC3 reference system, we estimated the galactocentric radii
of individual clusters. We adopted $(x_0,y_0)[pix]=(3692,1928)$ on the WFC3 frame as the centre of the galaxy.

From the centre of the galaxy (starting at 400 pixels $\sim350$ pc, i.e. avoiding the centre of the galaxy) until 
the location of the farthest cluster in our catalogue, we used radial bins with limits of $R[pc]=(372,2153,2700,3652,7155)$.
Each one of the bins encloses a total of 107 clusters, with the exception of the outer most  bin, where 108 objects
are in the bin. 

\subsubsection{Cluster formation rate (CFR)}
Our full cluster sample cover ages up to $\sim$300 Myr, however, the field stellar population (see Sect. 3.2) is constrained to ages less than 100 Myr due to incompleteness.  This provides the constraints on the age interval that can be used to estimate $\Gamma$, and reduces further our sample to a total of 278 star clusters, with a total of $[72,89,56,61]$ number of clusters per bin.
For each interval we assume that we are complete in detecting clusters more massive than $5\times10^3$ M$_{\odot}$. The lower limit of the cluster
studied was set to avoid problems coming from stochasticity in the initial mass function \citep[see e.g.][and references therein]{fouesneau10,silvavilla11}.
The total stellar mass which has formed in clusters is obtained summing the observed mass of each cluster in a given age bin 
and extrapolating the missing mass in clusters less massive than $5\times10^3$ M$_{\odot}$. 
To determine the total mass in clusters with mass between $5\times10^3$ M$_{\odot}$ and 100 M$_{\odot}$, we assume a power-law MF
with index -2 In Table \ref{tab:results} we report  the mean CFR values in each radial bin \citep[see][]{goddard10,adamo11}. 

\subsection{Star formation history (SFH)}

Following the methodology presented by \citet{silvavilla10}, we used the CMD method \citep{tosi91} to estimate the SFH inside the galaxy.
The CMD method uses the various observational uncertainties (photometric errors, distance, extinction) and theoretical isochrones to create a theoretical Hess
diagrams for different modelled SFHs. The observed and theoretical Hess diagrams are then compared and the best estimate of the SFH is obtained. To obtain the SFH of M83, we assumed the photometric
errors obtained in the photometric procedures above, $A_B$=0.29, a distance of $D\approx4.5$ Mpc \citep{thim03}, 
solar metallicity and theoretical Padova stellar evolutionary isochrones \citep{marigo08} obtained 
from their webpage, assuming a Chabrier IMF \citep{chabrier03}. The CMD method can take into account the binarity fraction and the mass ratio, 
but due to the lack of information on this parameter, we decided not to include binarity in our estimations. We refer the reader 
to \citet{silvavilla10} for more details on the CMD method and a set of various test performed
to estimate the reliability of the results.

\citet{silvavilla11} used the CMD method in two regions of the galaxy, with similar galactocentric radii as the HST/WFC3 pointings, and estimated
the SFH and star formation rates (SFR). The values presented by \citet{silvavilla11} are constrained to ages between 10 and 100 Myr. These limits
were chosen by the authors based on limits for completeness. Since the completeness limits for the WFC3 used in this work are significantly fainter than the previous ACS limits, we did not run additional completeness limits.  Instead, we adopted the more conservative completeness limits based on the ACS imaging. 

The SFH and SFR estimates were done over the WFC3 images, using the same radial bins used for the star clusters.
Due to the difference among the bands used by the HST images (i.e. F555W for the inner field and F547M for the outer field), for each bin
we divided the bin in "north" and "south" sample and use the respective set of isochrones in our codes. 
After each subregion for each bin has an estimated SFR, we summed them, obtaning the SFR for the bin.

The results obtained for each bin are presented in Table \ref{tab:results}.

\section{Discussion and Conclusions}

\begin{table*}
\caption{Radial variations of values for each bin used in this work.}
\begin{center}
\begin{tabular}{c c c c c c c c c c c c} \hline
Annuli & SFR & CFR & Area & $\Sigma_{SFR}\times10^{-3}$ & Q$_r$ & $\Omega^\dag$ & $\sigma_g^\dag$ & $\Sigma_g^\dag$ & $\Gamma_{theo}$ & $\Gamma_{obs}$  \\  
$[pc]$ & [M$_{\odot}$Yr$^{-1}$] & [M$_{\odot}$Yr$^{-1}$] & [Kpc$^{-2}$] & [M$_{\odot}$Yr$^{-1}$Kpc$^{-2}$] &  & [Kms$^{-1}$Kpc$^{-1}$] & [Kms$^{-1}$] & [M$_{\odot}$pc$^{-2}$] & [\%] & [\%]  \\ \hline \hline 
372 - 2153  & 0.36 & 0.05 & 8.76 & 41.1 & 3.1 & 90 & 25 & 75 & 37.2 & 14.7$\pm2.3$ \\
2153 - 2700  & 0.33 & 0.03 & 3.17 & 104.1 & 2.5 & 60 & 16 & 40 & 22.3 & 10.3$\pm3.1$ \\
2700 - 3652  & 0.34 & 0.01 & 7.02 & 48.4 & 2.4 & 50 & 14 & 30 & 18.4 & 3.8$\pm2.8$ \\
3652 - 7155  & 0.34 & 0.01 & 26.16 & 13.0 & 2.0 & 40 & 10 & 20 & 13.5 & 4.1$\pm2.6$ \\ \hline
\end{tabular}
\end{center}
\flushleft{$\dag$ Values taken from \citet{lundgren04a,lundgren04b}.The values are for the mean of each bin. }
\label{tab:results}
\end{table*}

\begin{figure*}
\includegraphics[scale=0.4]{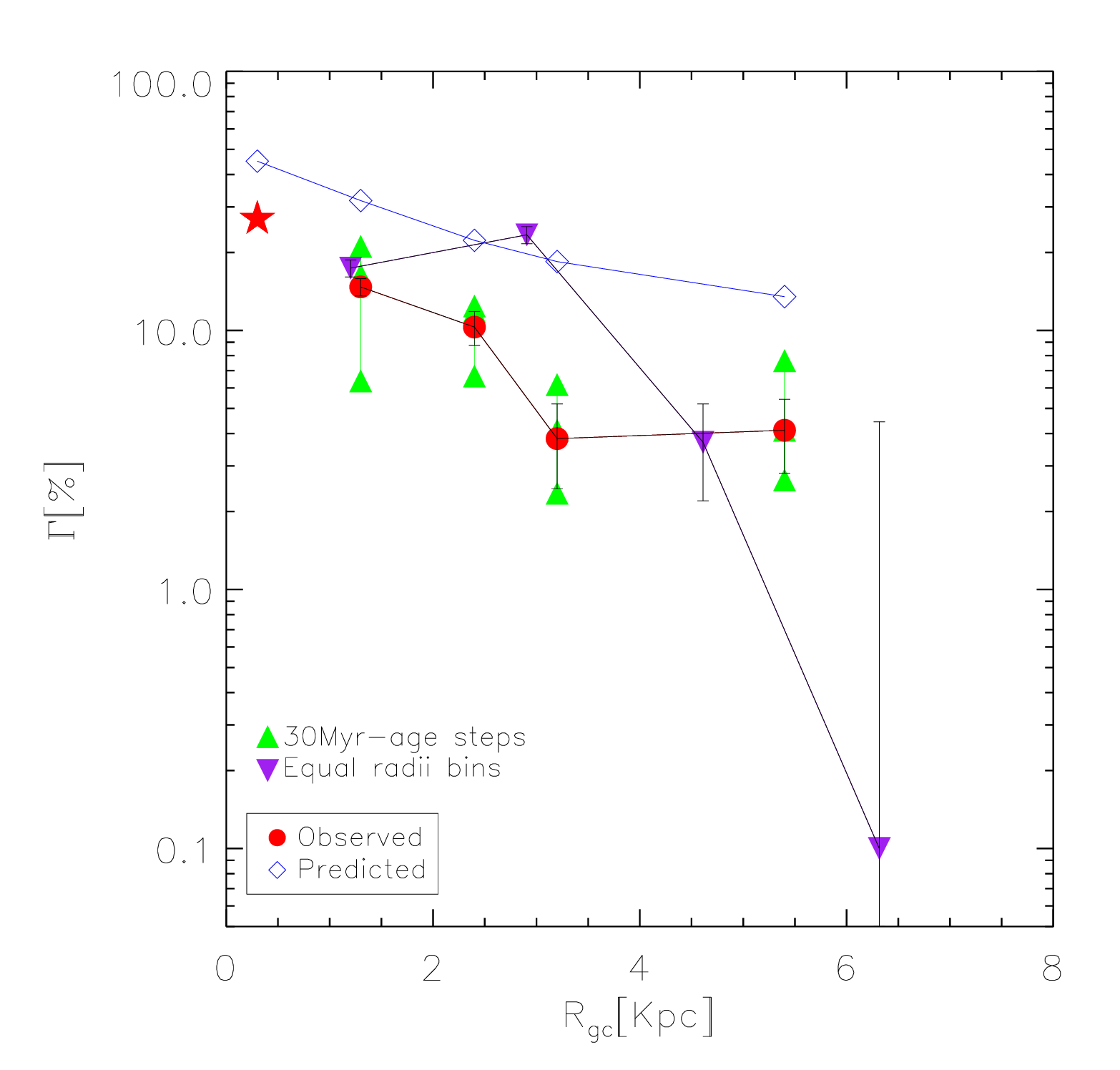}
\includegraphics[scale=0.405]{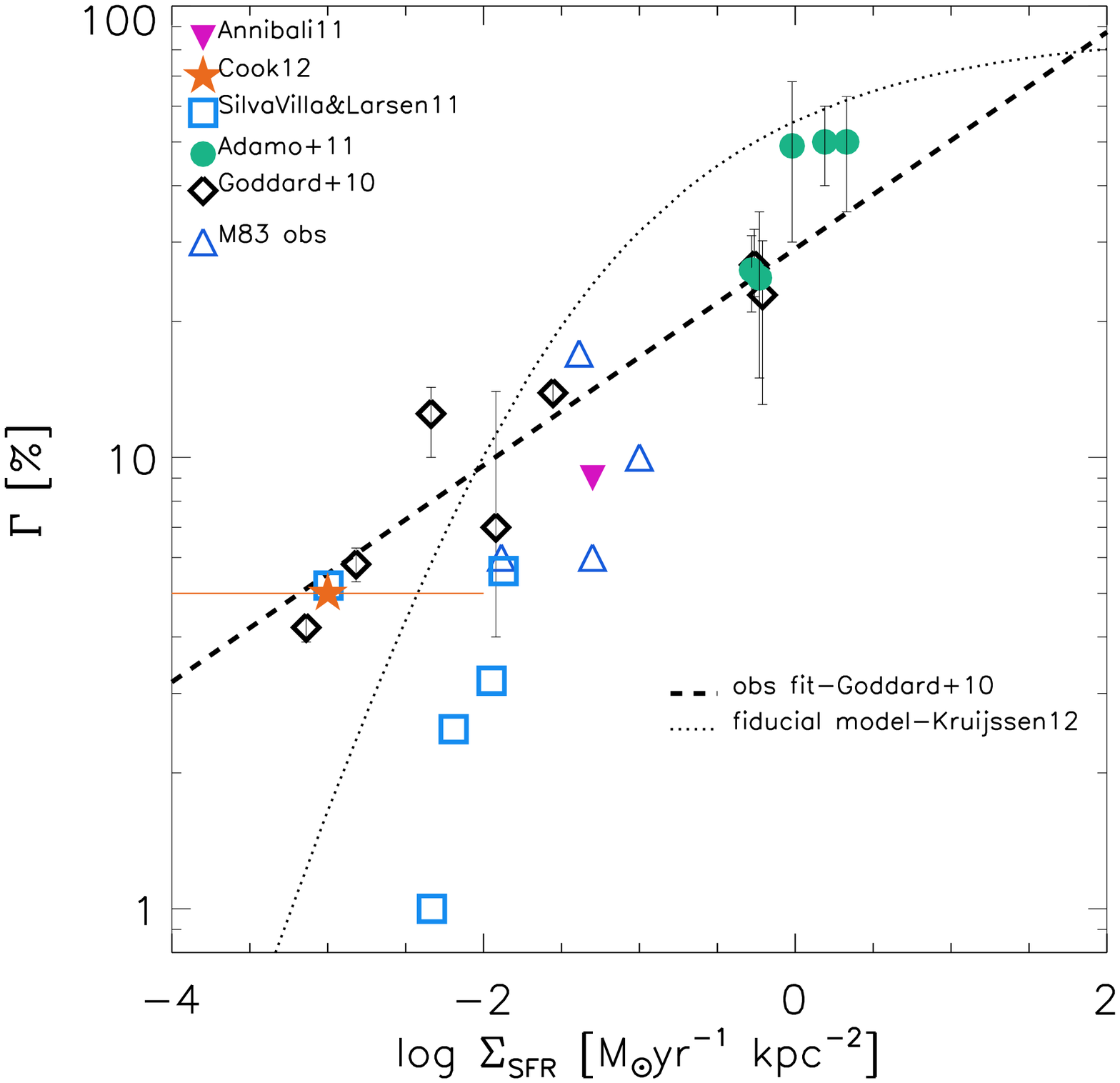}
\caption{{\em Left panel:} The radial variation of $\Gamma$ within M83. The filled (red) circles show the estimates for the four radial bins studied in this work, while the filled (red) star represents the estimated values for the central region of M83 \citep{goddard10}.  Open (blue) diamonds represent the theoretical expectations from the model of \citet{kruijssen12}.  While the predictions are a factor of $2-3$ too high, they predict the overall trend well. Green and purple triangles represent test using smaller age intervals inside each bin and equal radii bins, respectively. The errors are derived assuming a 0.3 dex in the age/mass estimations and 0.1 dex in the star formation history estimations. In the case of equal radii bins, the errors are larger at larger $R_{gc}$ due to small number of clusters.
{\em Right panel:} Comparison between the models and the observations for $\Sigma_{SFR}$ and $\Gamma$. Data taken from
\citet{goddard10,annibali11,adamo11,silvavilla11,cook12} as well as that presented here.}
\label{fig:gamma}
\end{figure*}

Based on the estimations obtained for the CFR and SFR (between 10-100Myr) in different locations of the galaxy, we can now investigate if, and how, $\Gamma$ varies within M83
(see Table \ref{tab:results}). This is shown in the left panel of Fig.~\ref{fig:gamma}. The errors associated to $\Gamma$ were derived assuming a 0.3 dex uncertainty on the estimated cluster age/mass and a 0.1 dex for the SFH.
There is a clear decreasing trend in the radial values of $\Gamma$ from inside out over the radii covered by our observations, showing values $\sim3$ times higher in the centre of the galaxy than in the outer regions.

We also investigated whether the observed decrease of $\Gamma$ as function of galactocentric distances is recovered using a different binning of the data. For this purpose, we divided the dataset using equally spaced radial bins. The new radii are located at the mean values of $R_{gc}=[1.2,2.9,4.6,6.3]$ Kpc. The estimations for the CFR were done in the same manner as for the variable binning cases, and we have assumed SFR=0.34 M$_{\odot}$yr$^{-1}$ for the 4 bins. We observed again a decrease of the values of $\Gamma$ with increasing $R_{gc}$, in good agreement with our previous results (see Fig. \ref{fig:gamma}, purple triangles). It is important to notice that the error bars are larger at large $R_{gc}$ due to the small number statistics.

Using age intervals of 30 Myr (between 10 and 100 Myr), we estimated $\Gamma$ values for three small age-bins inside each radial bin, as shown in Fig.~\ref{fig:gamma} (left panel, green triangles). These results suggest that there is a decrease of $\Gamma$ as a function not only of distance from the center of the galaxy, but also as a function of age (younger bins are on top of the older), most likely due to cluster disruption \citep{bastian11,bastian12}.

\citet{goddard10} estimated the CFR, SFR and $\Gamma$ for 
the central region of M83 (inside 300 pc) using the cluster catalogue from \citet{harris01} and the H$\alpha$ images
from \citet{meurer06}. The CFR is based on 45 massive clusters with ages between 3 to 10 Myr, which makes the
estimation different from ours. The SFR was derived by Goddard et al. using the H$\alpha$ images, 
assuming the transformation from \citet{kennicutt98}. The value obtained from Goddard et al. is shown as the filled (red) star in the left panel of Fig.~\ref{fig:gamma}, and follows the trend observed. 

We note that we have not corrected for cluster disruption in our analysis.  However, this is not expected to influence the trend observed in Fig.~\ref{fig:gamma}.  If cluster disruption is independent of mass/environment (e.g., \citealt{whitmore07}) then all points will be affected equally, and only the absolute value will be affected, not the overall trend.  If, however, cluster disruption does depend on mass/environment (e.g., \citealt{lamers05} - and found for M83 by \citealt{bastian12}) then the inner regions of the galaxy should have a shorter disruption time than the outer regions \citep{kruijssen11}.  This will destroy more clusters in the inner region, which in turn, will lower the inferred $\Gamma$ value in the inner region.  Hence, correcting for mass/environment dependent disruption would only exacerbate the observed trend.

We can compare our observed values and trend with the prediction of the framework presented by \citet{kruijssen12}.  The observational values used as input to estimate $\Gamma$ (using Kruijssen's IDL algorithm)
were taken from \citet[][i.e. gas surface density ($\Sigma_g$), velocity dispersion ($\sigma_g$), angular velocity 
($\Omega$). See Table \ref{tab:results}]{lundgren04a,lundgren04b}. 
We assumed the mean galacto-centric value for each one of the bins
and used equation 8 from \citet{kruijssen12} for the estimation of the \citep{toomre64} parameter $Q$.  The theoretical predictions are higher than our estimations although they predict the same trend as observed. Some part of this discrepancy comes from cluster disruption between the time of cluster formation and the adopted age interval, as is suggested by the green symbols in Figure 2. Any remaining discrepancy can easily be due to uncertain parameters in the cluster formation model, such as the adopted feedback efficiency (see Sect.~7.1 of \citealt{kruijssen12} for more details).
What is important to highlight is that both theory and observations agree on the
distribution, which is decreasing inside out across the galaxy at the rates observed. 

In order to compare with previous results, in the right panel of Fig. \ref{fig:gamma} we show the relation between $\Sigma_{SFR}$ and $\Gamma$ for the found bins discussed in the present work and previous estimates from the literature. The empirical fit from \citet{goddard10} as well as the fiducial model of \citet{kruijssen12} (dashed and dotted lines, respectively) as also shown.
The large scatter between model and observations can come from different sources 
(e.g. age- and mass ranges, source detections, incomplete samples, etc).

The observations presented here cannot differentiate between the fraction of stars that are originally bound in clusters and early disruption or mass loss (i.e. it is possible that all stars are born in clusters and that the fraction of clusters destroyed within the first 10 Myr is dependent on environment).  {\em However, for the first time, we have shown that the fraction of star formation that happens in long-lived bound clusters is tightly related to the local properties of the host galaxy, and specifically that it decreases with galactic radii.} 


\section*{Acknowledgments}
ES-V is a postdoctoral fellow supported by the Centre de Recherche en Astrophysique du Qu\'ebec (CRAQ).
NB was supported by a Royal Society University Research Fellowship.

\end{document}